\documentstyle[11pt,aaspp4,tighten]{article}

\tightenlines

\def\gsimeq
{\hbox{\raise0.5ex\hbox{$>\lower1.06ex\hbox{$\kern-1.07em{\sim}$}$}}}
\def\lsimeq
{\hbox{\raise0.5ex\hbox{$<\lower1.06ex\hbox{$\kern-1.07em{\sim}$}$}}}

\begin{document}

\title{The Optical Polarization and Warm Absorber in IRAS~17020+4544}

\author{Karen M. Leighly}
\affil{Columbia Astrophysics Laboratory, 538 West 120th Street, New
York, NY 10027, USA, leighly@ulisse.phys.columbia.edu}
\author{Laura E. Kay}
\affil{Department of Physics and Astronomy, Barnard College, Columbia
University, New York, NY 10027-6598}
\author{Beverley J. Wills, D. Wills, \& Dirk Grupe}
\affil{McDonald Observatory and Department of Astronomy, University of
Texas at Austin, TX 78712}

\slugcomment{Submitted to {\it The Astrophysical Journal Letters}}


\begin{abstract}

We report the detection of ionized absorption in the {\it ASCA}
spectrum of the narrow-line Seyfert 1 galaxy IRAS~17020+4544.  Subsequent
optical spectropolarimetry revealed high polarization increasing from
3\% in the red to 5\% in the blue, indicating electron or dust
scattering as a
likely origin.  The broad emission line H$\alpha$ is somewhat less
polarized than the continuum, supporting a location of the polarizing
material within the AGN.  The Balmer line decrement and reddened
optical spectrum supports the presence of a dusty warm absorber in
this object.
 
We compared the broad band optical polarization and ionized X-ray
absorption of a collection of Seyfert 1 and 1.5 galaxies, excluding
classes of objects that are likely to have significant neutral X-ray
absorption.  Warm absorber objects are generally more likely to have
high optical polarization than objects with no detected ionized
absorption.  This result lends additional support to the idea that the
warm absorber is associated with dust and implies either that dust
transmission is responsible for at least part of the 
polarization or that the
polarization is revealed because of the dimming of the optical spectrum.
Spectropolarimetry of Seyfert 1s generally locates the scattering
material inside the narrow-line region and often close to or within
the broad line region, consistent with estimates of the location of
the dusty warm absorber.

\end{abstract}

\newpage

\section{Introduction}

{\it ROSAT} and {\it ASCA} observations of Seyfert 1 nuclei produced
abundant evidence for highly ionized material in the line of sight.
Signatures of the ``warm absorber'' are present in the X-ray
spectra of about half of Seyfert 1 and 1.5s (Reynolds\markcite{32}
1997). Recently, it was noticed that the warm absorber is often
associated with optical reddening (Reynolds\markcite{32} 1997; Brandt,
Fabian \& Pounds\markcite{5} 1996), supporting the idea that the warm
absorber may coexist with dust.  Dust reddens the UV-optical continuum
spectrum, but it can also polarize it.  Thus one might expect high
optical polarization in Seyfert 1s that show warm absorber features
in their X-ray spectra.  As part of an archival analysis program of
{\it ASCA} data (Leighly et al.\markcite{20} 1997), we discovered
evidence for a warm absorber in the narrow-line Seyfert 1 galaxy
IRAS~17020+4544.  Noting that it also has a red optical spectrum, we
strongly suspected a dusty warm absorber would be present.  To test
the connection between warm absorbers and optical polarization, we
obtained spectropolarimetry and confirmed high polarization.

\section{Data and Analysis}

IRAS~17020+4544 is a member of the IRAS Point Source Catalog (IRAS
PSC) and was originally classified as a Seyfert 2 with redshift
z=0.0602 (De~Grijp et
al.\markcite{9} 1992).  The X-ray emission was first discovered in the
cross correlation between the {\it ROSAT} All Sky Survey and
the IRAS PSC (Boller et al.\markcite{4} 1992). Subsequent higher 
resolution spectroscopy revealed Fe~II lines and
H$\beta$ significantly broader than [O~III] (Moran, Halpern
\& Helfand\markcite{25} 1996), forcing reclassification as a 
Narrow-line Seyfert 1 galaxy
(NLS1; Osterbrock \& Pogge\markcite{29} 1985; Goodrich\markcite{14} 1989b).

\subsection{X-ray Spectral Analysis}

IRAS~17020+4544 was observed using {\it ASCA} on August 29 1995 and the
data were retrieved from the archive.  Standard analysis procedures
were followed (see e.g. Nandra et al.\markcite{27} 1997; Leighly et
al.\markcite{20} 1997 for details) resulting in approximately $33\,\rm
ks$ net exposure.  The source was bright, about $0.49\rm\, c/s$ in the
SIS0. Since our primary interest is in the warm absorber, we report only
analysis relevant to that here.

Preliminary fitting indicated spectral complexity in soft
X-rays.  We fitted a power law plus a narrow iron line to the data above
2~keV, then extrapolated that fit to lower energies, including a
neutral 
absorption column of $N_H=3.5\pm 0.5 \times 10^{20} \rm \,cm^{-2}$ measured
from the {\it ROSAT} PSPC spectrum; note that the Galactic column in
this direction is $2.2\times 10^{20}\rm\,cm^{-2}$ (Dickey \&
Lockman\markcite{10} 1990).  The residuals in Figure 1 show that the
source is absorbed in soft X-rays, and there is an edge near 0.7 keV
which is a characteristic signature of the warm absorber (e.g.
Reynolds\markcite{32} 1997).  A power law, narrow Gaussian (to model
the iron K$\alpha$ line) and 
additional absorption model over the whole range from 0.4--10.0~keV
gives a poor fit with $\chi^2$=1167 for 971 degrees of freedom
(d.o.f.).  Addition of an absorption edge improves the fit
significantly ($\Delta\chi^2=-131$ for two additional d.o.f.).  The
edge energy is $0.71\pm 0.02\rm\,keV $ (errors are 90\% for one
interesting parameter), roughly consistent with O~VII absorption.
Addition of another edge gives no improvement in fit.  Following
Reynolds\markcite{32} (1997) by fixing the two edge energies at
0.74 and 0.87~keV corresponding to O~VII and O~VIII results in a
slightly worse fit than the single edge fit by $\Delta\chi^2=8$ with
the optical depth of the O~VIII edge equal to zero.  This suggests
that the ionization state of the warm absorber is somewhat low
compared with objects studied by Reynolds\markcite{32} (1997).

\begin{figure}[t]
\vbox to2.5in{\rule{0pt}{2.5in}}
\includegraphics{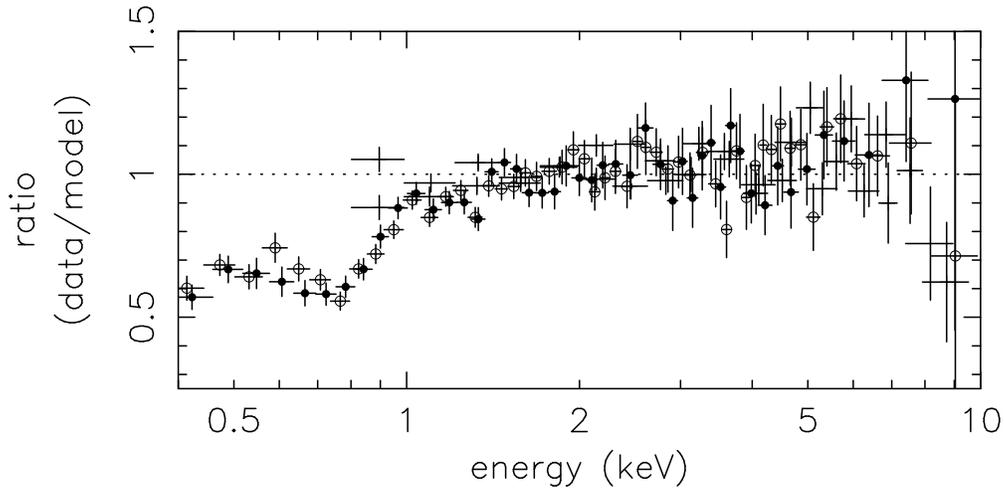}
\caption{The ratio of a power law plus absorption column obtained
from {\it ROSAT} spectrum ($3.5\times 10^{20}\rm\,cm^{-2}$; Galactic
$\rm N_H=2.2\times 10^{20}\rm\,cm^{-2}$) model fit above 2 keV and
extrapolated to low energies.  The open and solid circles correspond
to the SIS0 and SIS1 spectra, respectively, while the crosses
correspond to the GIS2 and GIS3 spectra.}
\end{figure}

Next we model the warm absorber with the photoionization model {\it
absori} available in XSPEC (Magdziarz \& Zdziarski\markcite{21} 1995).
This model results in a somewhat poorer fit ($\chi^2$=1071/969 d.o.f.)
than the edge model but provides an estimate of the ionized column
density of $N_{w}=2.5^{+0.6}_{-0.5}\times 10^{21}\rm \,cm^{-2}$ and
the ionization parameter $\xi=9.5^{+19}_{-7.6}$.  Evidence for
emission around 1 keV remains in the residuals. This feature can be
modeled as a marginally resolved line at 1.1 keV with equivalent width
75~eV, and then $N_w=3.5^{+0.6}_{-0.5} \times 10^{21} \rm \, cm^{-2}$,
$\xi=6.1^{+9.8}_{-3.3}$, intrinsic $N_H=5.5^{+2.1}_{-1.8}\times
10^{20}\rm\, cm^{-2}$ and $\chi^2=1016/967$~d.o.f.  The 1~keV feature
may be similar to the emission features seen in the X-ray spectrum of
other NLS1s, possibly a blend of photoionized iron and neon emission
lines (e.g.  PG~1244+026: Fiore et al.\markcite{11} 1997; Ton~S180,
Akn~564: Leighly et al.\markcite{20} 1997). 

\subsection{Spectropolarimetry}

We obtained
spectropolarimetry data on IRAS~17020+4544 at the Lick Observatory 3m
telescope with the KAST spectrograph (e.g. Martel\markcite{22}~1996)
and at the McDonald Observatory 2.7m telescope with the Large
Cassegrain Spectrograph (e.g. Hines \& Wills\markcite{45}~1993).
Figure 2 shows spectropolarimetry results. 
The polarization position angle is constant at about $166^\circ$
and is therefore not shown.  We measured the host galaxy axial ratio
on the Digitized Sky Survey image to be $0.55\pm 0.02$ with position
angle $168\pm 1^\circ$, in good agreement with the polarization
position angle.  Broad-band imaging polarimetry measurements with the
McDonald Observatory 2.1-m telescope (Grupe et al.\markcite{17}~1997) agree in
position angle and in the blue but find lower polarization in the red,
probably a result of a larger aperture ($7.4^{\prime\prime}$ diameter compared with 
$2^{\prime\prime}$ slits.)  We find (filter--effective wavelength--percentage polarization):
U--3600\AA--$7.1\pm 3.1$; CuSO4--4200\AA--$3.93\pm0.35$;
none--5700\AA--$2.42\pm 0.18$; RG~630--7600\AA--$1.68\pm 0.20$). The
spectropolarimetry data indicate that the continuum is polarized at
about 3\% at the red end increasing to 5\% at the blue end.  The
Balmer lines of H$\alpha$ (and perhaps H$\beta$) are less polarized
than the continuum average, and the [NII] $\lambda\lambda$ 6548,6583 
and [OIII] $\lambda\lambda$ 4959,5007 lines may be
slightly less polarized than the Balmer lines.  In polarized flux the
[OIII]/H$\beta$ ratio is lower than in direct flux and the
[NII]$\lambda$6583/H$\alpha$ ratio may also be slightly lower.  The
direct and polarized flux widths are reasonably similar in H$\alpha$;
this is probably true but more difficult to measure in H$\beta$.
Since the widths of the Balmer lines are similar in polarized and
direct flux, and the continuum polarization clearly rises to the blue,
we can conclude that reflection by dust or electrons is a likely cause of the
polarization, although dust transmission may contribute (see below).
This is in agreement with a sample of NLS1s observed by
Goodrich\markcite{14}~(1989b).  We note also that high polarization is
often found in dusty IRAS-selected AGN (e.g. Wills \& Hines 1997).

In direct flux, H$\alpha$/H$\beta=8.4$.  Following Reynolds et
al.\markcite{33}~1997, for a Galactic interstellar medium dust to gas
ratio and assuming intrinsic ratio of 3.1 (Veilleux \&
Osterbrock\markcite{42}~1987), we derive $4.0\times 10^{21}\rm
cm^{-2}$ for the column density, roughly consistent with the ionized
column density measured in the X-rays.  This result supports the
association of the dust with the ionized gas, and suggests that the
broad lines are seen through most of the obscuring screen.  In
polarized flux H$\alpha$/H$\beta$ was $\sim$3.5; however, statistics
were not good enough to measure the lines accurately.  The polarized
flux spectrum is nearly flat, as the reddening seen in the direct flux
spectrum cancels the rise to the blue in $P$.

\section{Discussion}

Based on the discovery of the warm absorber in IRAS~17020+4544, we
postulated high polarization and found that it was present.  We
collected data from the literature to test the generality of the
association between the presence of the warm absorber and high optical
polarization.

\begin{figure}[b]
\vbox to5.5in{\rule{0pt}{5.5in}}
\includegraphics{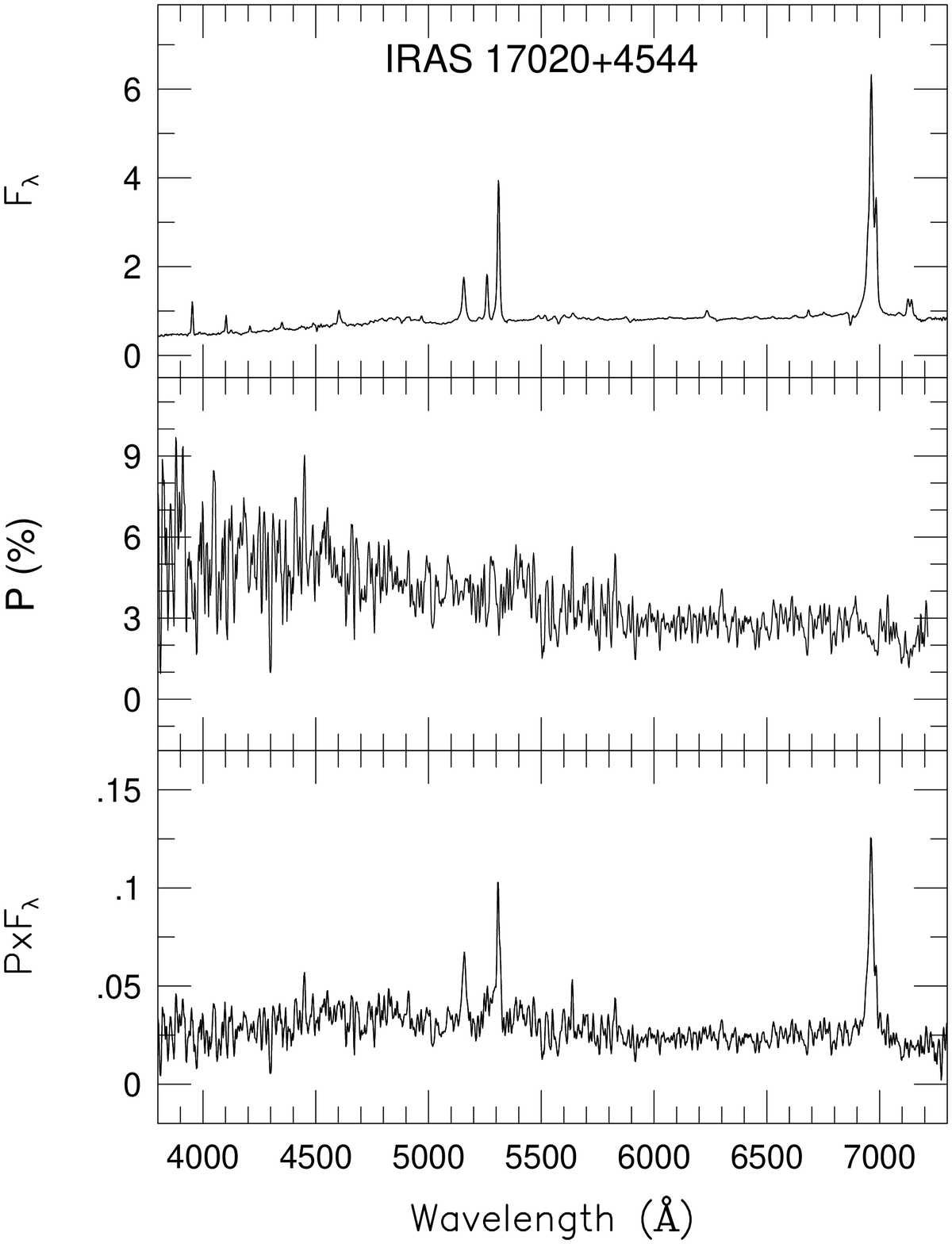}
\caption{Spectropolarimetry of IRAS 17020+4544
uncorrected for reddening or redshift. 
The Lick data (60 minute exposure on 1997 March 8) were matched to
the McDonald data (200 minutes  exposure on 1997 April 8 and 10), taking
into account small differences in polarization in the red, possibly the result
of host galaxy starlight. 
The flux is in units of $10^{-15} \rm erg/sec/cm^{2}/$\AA.
The second panel is the polarization (strictly Stokes q rotated to 
position angle 163 degrees), and the third is the corresponding polarized
flux.  No other corrections were made, as lack of strong
stellar absorption lines in the direct flux spectrum suggests little 
contamination.}
\end{figure}

The sample was chosen carefully.  Because our goal was to test the
association of the {\it ionized} absorber with optical polarization,
we excluded objects in which high {\it neutral} columns are expected,
 since dust associated with the neutral column could
also produce polarization.  Therefore, we included Seyfert 1s,
1.5s and narrow-line Seyfert 1s, but excluded Seyfert 1.8s, 1.9s and
NELGs, which are often  reddened, suffer X-ray
absorption and lie in galaxies viewed at high inclination
angle (Goodrich\markcite{13}\markcite{15} 1989a, 1995; Lawrence \&
Elvis\markcite{18} 1982; Mushotzky\markcite{26} 1982; Forster et
al.\markcite{12} 1997).  We also excluded objects at low Galactic
latitude to avoid polarization by the Galactic interstellar medium;
nevertheless this contributes a systematic error of about 0.3--0.4\%.
Optical polarization of Seyfert 1s is  correlated with
the axial ratio of the host galaxy (Berriman\markcite{2} 1989; Thompson \&
Martin\markcite{38} 1988) so we exclude objects with low $b/a$
(IC~4329a, Mrk~1040) unless differences in line and continuum
polarization indicates the absorber is inside the AGN (3A~0557-383;
Brindle et al.\markcite{8} 1990b).  Broad-line radio galaxies are also
excluded since a contribution to their polarization may come 
from a nonthermal component (Rudy et al.\markcite{35} 1983;
Antonucci\markcite{1} 1984), and they also sometimes show weak
intrinsic neutral X-ray absorption (Woz\'niak et al.\markcite{41}
1997).

The resulting sample comprised all the objects from
Reynolds\markcite{32}~(1997) excluding those listed above,
MR~2251-178, for which we found no polarization measurement, and
3C~273.  We included also Mrk~766 (Leighly et al.\markcite{19} 1996),
IRAS~13349+2438 (Brandt et al.\markcite{6}~1997), 3A~0557-385 (Turner,
Netzer \& George\markcite{39}~1996), NGC~7213 (Otani\markcite{30}
1996) and Akn~120, I~Zw~1, Mrk~478, Mrk~279, Mrk~110 (from the archive
and analyzed by KML following Section~2.1 and Reynolds\markcite{32}
1997; warm absorbers were not detected in these objects). Broad-band
polarization measurements in the band 3800-5600\AA\, which is blue
enough to avoid dilution by cool starlight, were used.
 The values were predominantly taken from
Berriman\markcite{2} (1989), except for Mrk~766 (Goodrich\markcite{14}
1989b), IRAS~13349+2438 (average of B
\& V band, Wills et al.\markcite{40} 1992), Mrk~335 and Mrk~110
(Berriman et al.\markcite{3} 1990), and IRAS~17020+4544, presented
here.

The polarization versus column density is shown in Figure 3.  Open
symbols mark the ionized absorber column density in objects in which a 
warm absorber was detected, while solid symbols plot the excess
neutral column density over Galactic in objects with no detectable
warm absorbers. Objects with no
measurable excess neutral column are assigned $N_H=1
\times 10^{20}\rm\,cm^{-2}$, approximately the level of systematic
error  from {\it ASCA} spectra, and errors equal to max\{fit upper
limit, $1\times 10^{20}\rm\,cm^{-2}$\}.
The neutral column density of the warm absorber objects is not taken
into account, since the warm column density is very much larger than
the cold column density in all cases except 3A~0557-383.  This plot
shows that objects with high optical polarization ($\gsimeq 1$\%) are
very likely to have warm absorbers.  However, the converse is not
generally true; i.e.  objects with high ionized columns don't necessarily
have high optical polarization.  Two notable examples, NGC~3783 and
NGC~3516, are discussed below. Note that scatter is expected since the
warm absorber can in principle respond rapidly to ionizing flux
changes while dust properties are expected to change on much longer
time scales.  Nevertheless, the KS test indicates a different
distribution of polarization of warm and cold absorber objects with
99\% confidence.

\begin{figure}[t]
\vbox to2.75in{\rule{0pt}{2.75in}}
\includegraphics{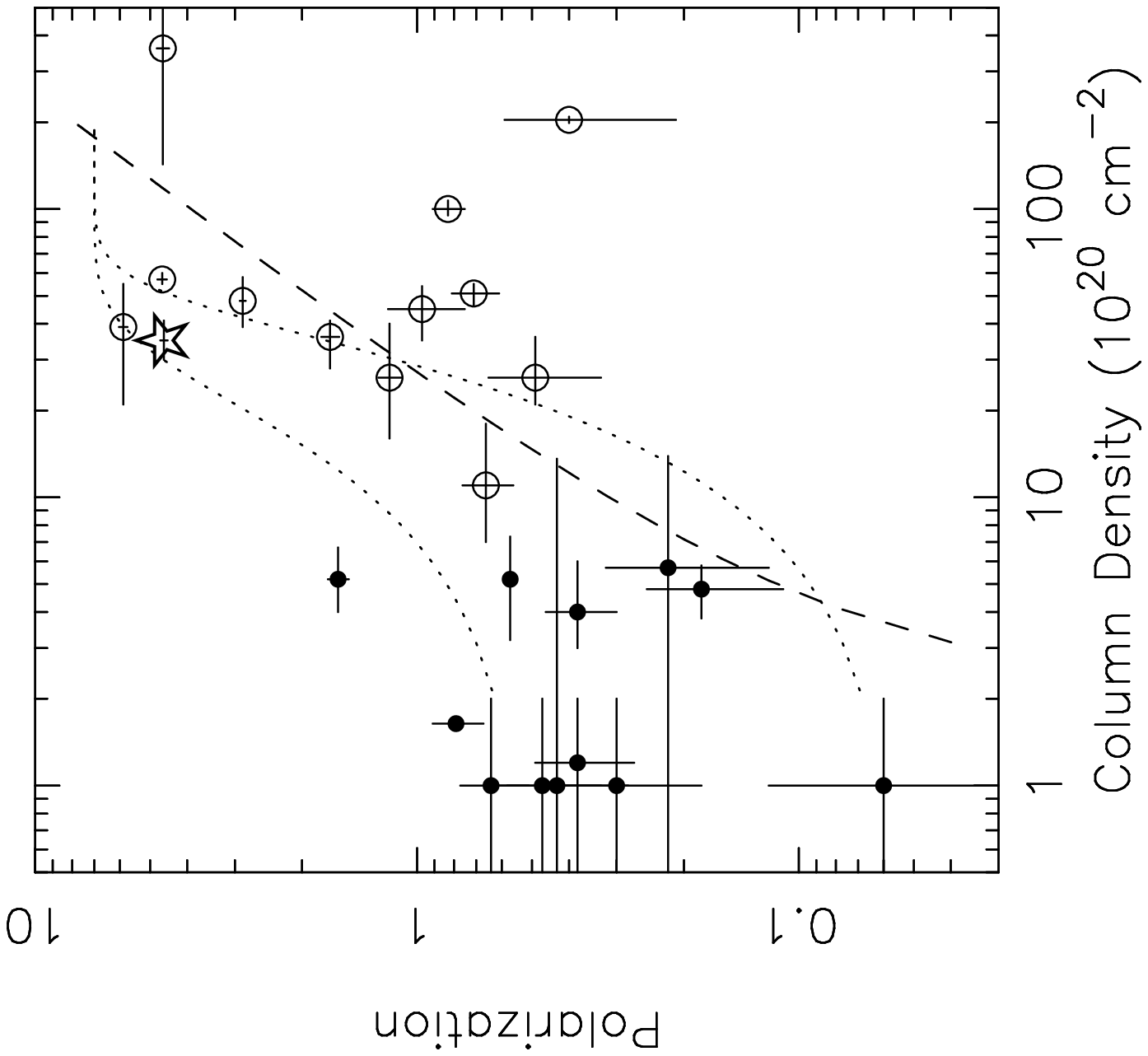}
\caption{Absorption column versus polarization.  Open symbols:
ionized column density of objects with warm absorbers detected in {\it
ASCA} spectra; solid symbols: excess neutral absorption over Galactic
in objects with no detectable warm absorber (see text). The star 
marks the position of
IRAS~17020+4544.  Dashed curve: predicted maximum polarization from 
dust transmission.  Dotted curves: predicted polarization from
scattering assuming 7\% intrinsic polarization  and two assumed
ratios of scattered to direct flux: 0.1 and 0.01 for upper and lower
curves, respectively. }
\end{figure}

Brandt, Fabian \& Pounds\markcite{5} (1996) first speculated that the
high degree of optical reddening in IRAS~13349+2438 could be
reconciled with the bright soft X-ray emission only if the gas
associated with the dust responsible for reddening were ionized.
Reynolds\markcite{32} (1997) found a strong relationship between the
reddening of the optical spectrum and the optical depth of the ionized
absorber O~VII edge in a sample of 24 AGN.  In a sample of bright soft
X-ray selected objects, Grupe et al.\markcite{17}~(1997) found that
significant polarization occurred in objects in which the degree of
optical reddening was too high to be consistent with the relatively
unabsorbed soft X-ray spectra unless the gas were ionized.  These results
all support the association of the warm absorber with dust.

Generally speaking, in Seyfert 1s and 1.5s, the narrow emission lines
are less polarized than the broad lines which are in turn less
polarized than the continuum, indicating that the scattering material
is interior to the narrow line region (NLR).  Many Seyfert 1s and 1.5s show
changes in position angle across the broad emission lines (e.g. Goodrich \& Miller\markcite{16} 1994; Martel\markcite{22}
1996), and variability in the polarization properties on time scales
of months to years (Martel\markcite{22}
1996; Smith et al.\markcite{37}~1997), indicating
that the scattering region is not much larger than the broad line
region (BLR).   If the dust is associated with
the warm absorber, it must be located far enough from the nucleus that 
the dust does not evaporate.  MCG--6-30-15 apparently
has an inner and outer warm absorber (Otani et al.\markcite{31}~1996), 
and it is probable that the outer one is dusty 
(Reynolds et al.\markcite{33} 1997).  Since it is unlikely that the
dust could condense out of ionized gas, a source of dusty gas which
can then be ionized is required; this could be a wind off the
molecular torus lying at a radius between the BLR and NLR radii in unified models (Reynolds et al.\markcite{33}~1997).
It has been suggested that dust absorption may naturally result in a
line free zone between the BLR and NLR
(Netzer \& Laor\markcite{28}~1993). The narrow ``associated'' UV
absorption lines may originate in warm absorber material (e.g. Mathur,
Elvis \& Wilkes\markcite{23}~1995); since these are superimposed on
the broad emission lines, a location outside the BLR is
required.  These results all support a
similar location for the polarizing material and the dusty warm
absorber.  

The warm absorber measures conditions in line of sight gas.  If the
same material is responsible for the warm absorber, reddening and
polarization, then dust transmission must be responsible for at least
part of the polarization.  
The dashed line in Fig.\ 3 shows the predicted polarization versus
column density for the dust transmission mechanism.  We assumed the
empirical laws appropriate for the Galactic interstellar medium
(Clayton \& Cardelli\markcite{44}~1988), including maximum
polarization $P_{max}=9$\%$\,$E$(B-V)$, ratio of total-to-selective
extinction $R_V=3.1$, and the $\rm N_H$-to-reddening relation from 
Heiles, Kulkarni \& Stark\markcite{43}~1981.  Polarization
should lie below this line if dust transmission is the only
polarizing mechanism.  However, dust and electron scattering may also
contribute  to the polarization; furthermore, the geometry,
grain alignment, dust composition, and dust-to-gas ratio are probably
different in the AGN.
Another possibility is that high polarization is revealed as a
consequence of the suppression of the unpolarized direct (rather than
scattered) light by reddening (e.g.  Wills et al.\markcite{40}~1992).
The dotted lines in Fig.\ 3 show the predicted polarization versus
$N_H$ for this model.  We assumed that the intrinsic maximum
polarization is 7\%, seen when the direct continuum is completely
attenuated, and that the ratio of scattered to direct light is 0.1 or
0.01 (upper and lower curves, respectively).

While NGC~3516 and NGC~3783 have among the highest ionized column
densities ($N_w=100$ and $204 \times 10^{20}\,\rm cm^{-2}$), they have
only moderate polarization ($\sim 0.45$\% and $\sim 0.80$\%, respectively)
and are also not substantially reddened
(Reynolds\markcite{32} 1997).  Perhaps only the inner warm absorber is
present or the dust has been destroyed in these objects. Electron
scattering could be the origin of the moderate polarization.  No
strong wavelength dependence is seen in either object, consistent with
this idea (NGC~3516: Martel\markcite{22}~1996; NGC~3783: Brindle et
al.\markcite{7}~1990a,b).  Goodrich \& Miller\markcite{16}~(1994) find
that a maximum of 7\% polarization can be
obtained in Seyfert 1s when the optical depth $\tau=1$.  The
ionized column densities present an optical depth of $\tau=0.1$ and
therefore the observed polarization could be consistent with an origin
of free electrons in the ionized gas.

It would be interesting to extend this work to include Seyferts with
significant neutral absorption: Seyfert 1.8s, 1.9s, and NELGs.
All together, this
may support a picture in which the dust, ionized gas and broad
emission line clouds have a common origin in Seyfert 1s and
intermediate type Seyferts, with decreasing inclination angle reducing
the amount of obscuring material in the line-of-sight, but revealing
gas of increasing ionization parameter.

\acknowledgements

The authors thank Ross Cohen for contributing the Lick observing time.
KML thanks R. Mushotzky for useful comments on a draft.  KML, LEK,  BJW and
DG gratefully acknowledge support through NAG5-3307 ({\it ASCA}), NSF
Career grant Ast 9501835, GO-06766 (STScI) and NAG5-3431 (LTSA).


\clearpage

\begin{references}

\reference{1} Antonucci, R. R. J. 1984, \apj, 278, 499

\reference{2} Berriman, G., 1989, \apj, 345, 713

\reference{3} Berriman, G., Schmidt, G. D., West, S. C., \& Stockman,
H. S. 1990, ApJS, 74, 869

\reference{4} Boller, Th., Meurs, E. J. A., Brinkmann, W., Fink, H.,
Zimmerman, U., \& Adorf, H.-M., 1992, A\&A, 261, 57

\reference{5} Brandt, W. N., Fabian, A. C., \& Pounds, K. A., 1996,
\mnras, 278, 326

\reference{6} Brandt, W. N., Mathur, S., Reynolds, C. S., \& Elvis,
M., 1997, \mnras, in press

\reference{7} Brindle, C., Hough, J. H., Bailey, J. A., Axon, D. J.,
Ward, M. J., Sparks, W. B., \& McLean, I. S., 1990a, \mnras, 244, 577

\reference{8} Brindle, C., Hough, J. H., Bailey, J. A., Axon, D. J.,
Ward, M. J., Sparks, W. B., \& McLean, I. S., 1990b, \mnras, 244, 604 

\reference{44} Clayton, G. C., \& Cardelli, J. A., 1988, \aj, 95, 695

\reference{9} De~Grijp, M. H. K., Keel, W. C., Miley, G. K.,
Goudfrooij, P., \& Lub, J. 1992, A\&AS, 96, 389

\reference{10} Dickey, J. M., \& Lockman, F. J., 1990, AAR\&A, 28, 215

\reference{11} Fiore, F., et al. 1997, in prep.

\reference{12} Forster, K., et al. 1997, in prep.

\reference{13} Goodrich, R. W. 1989a, \apj, 340, 190

\reference{14} Goodrich, R. W. 1989b, \apj, 342, 224

\reference{15} Goodrich, R. W. 1995, \apj, 440, 141

\reference{16} Goodrich, R. W., \& Miller, J. S., 1994, \apj, 434, 82

\reference{17} Grupe, D., Wills, B. J., Wills, D., \& Beuermann, K.,
1997, A\&A, submitted

\reference{43} Heiles, C., Kulkarni, S., \& Stark, A. A. 1981, ApJL,
247, 73

\reference{45} Hines, D.C., \& Wills, B.J. 1993, \apj, 415, 82

\reference{18} Lawrence, A., \& Elvis, M. 1982, \apj, 256, 410

\reference{19} Leighly, K. M., Mushotzky, R. F., Yaqoob, T., Kunieda,
H., \& Edelson, R., 1996, \apj, 469, 147

\reference{20} Leighly, K. M., et al. 1997 in preparation

\reference{21} Magdziarz, P., \& Zdziarski, A. A. 1995, \mnras, 273,
837 

\reference{22} Martel, A. R. 1996, PhD thesis, University of
California Santa Cruz

\reference{23} Mathur, S., Elvis, M., \& Wilkes, B., 1995, \apj, 452, 230

\reference{24} Miller, J.S., Robinson, L. B., and Goodrich, R. W. 1988, in
{\it Instrumentation for Ground Based Astronomy}, ed. L. B. Robinson (New York:
Springer) p 157. 

\reference{25} Moran, E. C., Halpern, J. P., \& Helfand, D. J. 1996,
ApJS, 106, 341

\reference{26} Mushotzky, R. F. 1982, \apj, 256, 92

\reference{27} Nandra, K., et al.  George, I. M., Mushotzky, R. F., Turner, T.
J., \& Yaqoob, T. 1997, \apj, 476, 70

\reference{28} Netzer, H., \& Laor, A., 1993, ApJL, 404, 51

\reference{29} Osterbrock, D. E. \& Pogge, R. W. 1985, \apj, 297, 166

\reference{30} Otani, C. 1996, PhD Thesis, Tokyo University

\reference{31} Otani, C., et al. 1996, PASJ, 48, 211 

\reference{32} Reynolds, C. S., 1997, \mnras, 286, 513

\reference{33} Reynolds, C. S., Ward, M. J., Fabian, A. C., \&
Celotti, A., 1997, \apj\ in press

\reference{35} Rudy, R. J., Schmidt, G. D., Stockman, H. S., \&
Moore, R. L., 1983, \apj, 271, 59

\reference{37} Smith, P. S., Schmidt, G. D., Allen, R. G., \& Hines,
D. J., 1997, \apj\ in press

\reference{38} Thompson, I. A., \& Martin, P. G., 1988, \apj, 330, 121

\reference{39} Turner, T. J., Netzer, H., \& George, I. M. 1996, \apj,
463, 134

\reference{42} Veilleux, S., \& Osterbrock, D. E., 1987, ApJS, 63, 295

\reference{36} Wills, B. J., \& Hines, D. C., 1997, in proc. ``Mass
Ejection in AGN'', in press

\reference{40} Wills, B. J., Wills, D., Evans, N. J., Natta, A.,
Thompson, K. L., Breger, M., \& Sitko, M. L. 1992, \apj, 400, 96

\reference{41} Woz\'niak, P. R., Zdziarski, A. A., Smith, D.,
Madejski, G. M., \& Johnson, W. N., 1997, \mnras\ in press

\end{references}
\end{document}